\documentclass[10pt,letterpapers,floats]{article}
\usepackage{amsfonts,color,epsfig}
\usepackage{times}
\usepackage{multirow}
\usepackage[multiple]{footmisc}
\newcommand{\nc}{\newcommand}

\newcommand{\rnc}{\renewcommand}
\renewcommand{\thefootnote}{\fnsymbol{footnote}}
\makeatletter
\rnc{\theequation}{\thesection.\arabic{equation}}
\@addtoreset{equation}{section}
\makeatother
\nc{\fig}[5] {
\begin{figure}[!htbp]
    \begin{center}
    \leavevmode
    \centerline{
        \includegraphics[width=#1, height=#2]{#3}
        }
    \caption[]{#4}
    \label{#5}
    \end{center}
\end{figure}}
\nc{\figs}[8]{
\begin{figure}[!htbp]
    \begin{center}
    \leavevmode
    \centerline{
        \includegraphics[width=#1, height=#2]{#3}
        \includegraphics[width=#4, height=#5]{#6}
        }
    \caption[]{#7}
    \label{#8}
    \end{center}
\end{figure}}

\begin{document}
\begin{flushright}
{\small \tt arXiv:0906.4428\\
\tt [gr-qc]}
\end{flushright}
\vspace{5mm}
\begin{center}
{\Large {\bf Gravitational Collapse of the Shells with the Smeared Gravitational Source in
Noncommutative Geometry}}\\[10mm]
{John J. Oh$^a$\footnote{Email: johnoh@nims.re.kr} and
Chanyong Park$^b$\footnote{Email: cyong21@sogang.ac.kr}}\\[10mm]
{\small ${}^a$ {\it Division of Computational Sciences in Mathematics,\\
National Institute for Mathematical Sciences,\\ Daejeon 305-340, South Korea\\[0pt]
${}^{b}$ \it Center for Quantum Space-Time, Sogang University, \\Seoul 121-742, South Korea
\\[0pt]
}}
\end{center}
\vspace{2mm}
\begin{abstract}
We study the formation of the (noncommutative) Schwarzschild black hole from
{collapsing shell {of the} generalized matters {containing} polytropic and Chaplygin gas}.
We show that
this collapsing shell depending on various parameters forms either
a black hole or a naked singular shell with the help
of the pressure.
Furthermore, by considering the smeared gravitational sources, we investigate the
noncommutative black holes formation. Though this {mild} noncommutative correction of matters
cannot ultimately
resolve the emergence of the naked singularity, we show that in some parameter region
the collapsing shell evolves to a noncommutative black hole before becoming a naked singular
shell.
\end{abstract}
\vspace{3mm}


\vspace{5mm}

\hspace{10.5cm}{Typeset Using \LaTeX}
\newpage
\renewcommand{\thefootnote}{\arabic{footnote}}
\setcounter{footnote}{0}

\section{Introduction} \label{sec:intro}

One of the fascinating topics in general relativity is the dynamical formation of black holes
from gravitationally collapsing matter.  A remarkable breakthrough that emerged from such
studies is mathematical theorem \cite{sing} dubbed ``singularity theorem''.
According to the theorem, if a trapped surface forms during the collapse of physically
reasonable matter, the resultant spacetime geometry inevitably leads to the creation of
a singularity, assuming that there are no closed timelike curves (CTC{s}). This fact,
however, does not imply that the collapse process is necessarily accompanied with the
formation of black holes since there is no reason that the trapped surface hides the
singularity from distant observers. Thus it has been suggested that a black hole must
be formed from gravitationally collapsing physical matter with generic initial data in
order to preserve future predictability beyond  singularity formation, at least for distant
observers. This is called the ``Cosmic Censorship Hypothesis'' \cite{cch,cch2}. The issue
of whether or not this hypothesis is true remains {unsolved} in the context of the general
relativity -- there are some counter-examples against the hypothesis, producing a naked
singularity \cite{counter}.

On the other hand, in the three-dimensional Einstein's gravity with a cosmological constant,
it has been shown that the cosmic censorship violation happens under a broad variety of
initial conditions from the gravitationally collapsing shells of diverse static
\cite{Mann:2006yu} and the rotating \cite{Mann:2008rx} matter configurations. As the
shell collapses its energy density (and pressure, if any) diverges in finite proper time.
The interesting point is that even if the exterior space develops no curvature singularity
(since the exterior three-dimensional spacetime has constant curvature), the energy-momentum
tensor of the shell diverges in finite proper time, where the time-evolution of the shell's
equation of motion breaks down. As a consequence, this leads to a violation of  cosmic
censorship in that there is no definite way to evolve the equation of motion beyond this
point and a Cauchy horizon appears if the singularity is not screened by an event horizon.
This behavior also appears in the previous studies of the gravitationally collapsing dust
ball in
three-dimensions \cite{mr} and the topological black hole formation in four dimensions
\cite{ms,Mann:1997jb, Mann:1997iz}.

As seen in previous studies, dealing with a singularity might cause some troubles in studying
physics in the
context of, at least, general relativity. This is due to the fact that the theory of gravity
we study is
adhere to the Riemmanian geometry that has a drawback in handling a singular point. This is
{the reason}
why the quantum gravity is strongly required. There are two different prospects toward quantum
gravity so
far -- string theory and loop quantum gravity, in both which there should be somewhat
noncommutative
modification of spacetime geometry that possesses a symplectic structure.
In this sense, such a modification
of gravity might resolve the {singularity} problem.

In this paper, we study the formation of Schwarzschild black hole from gravitationally
collapsing shells of
polytropic matter and Chaplygin gas. And we study the formation of noncommutative
Schwarzschild black hole
slightly modified by the smeared gravitational source and
{classify the possible scenarios for the black hole ( or naked singular shell ) formation
in the various parameter regions.}
In Sec. \ref{sec:shell}, we present a shell formalism and an introduction
of the smeared gravitational source that yields noncommutative Schwarzschild black
hole. In Sec. \ref{sec:SS},
we investigate gravitational collapse scenarios for collapsing polytropic and the
Chaplygin gas shells,
in which it can form either a black hole or a naked singularity, depending on its
initial data.
In Sec. \ref{sec:NCSS}, we investigate the formation of {the} noncommutative black
hole from the
collapsing shells, where new-type of matter (non-polytrope type)
is needed to solve the junction equation. Finally, we shall summarize
and discuss the results in Sec. \ref{sec:discuss}.

\section{Shell Collapse and Smeared Gravitational Source}\label{sec:shell}

\subsection{Hypersurface Formalism with Smeared Source}

Let us consider a hypersurface in four-dimensional manifolds which is described by a
surface stress-energy tensor denoted by ${\mathcal S}_{\mu\nu}$. Then the manifold is divided
into two parts -- exterior and interior regions. By introducing {the} Heaviside distribution
function ${\mathcal H}(\sigma)$\footnote{${\mathcal H}(\sigma)$ is $1$ if $\sigma>0$, $0$
if $\sigma <0$, and indeterminate if $\sigma=0$. Its crucial properties are
${\mathcal H}^2(\sigma) = {\mathcal H}(\sigma)$,
${\mathcal H}(\sigma){\mathcal H}(-\sigma)=0$,
and $d{\mathcal H}(\sigma)/d\sigma = \delta(\sigma)$,
where $\delta(\sigma)$ is a delta function.},
we have $g_{\mu\nu}={\mathcal H}(\sigma) g_{\mu\nu}^{+} +{\mathcal H}(-\sigma)g_{\mu\nu}^{-}$,
where $\sigma$ is a geodesic coordinate. Differentiating this yields a smoothness condition
for the metric, $[g_{\mu\nu}]=0$, where $[A]=A^{+}-A^{-}$, which is called the first
junction condition.

Computing the Riemann tensor leads to the Einstein tensors in both regions and the additional
term from the shell's edge
\begin{equation}
G_{\mu\nu} = {\mathcal H}(\sigma) G_{\mu\nu}^{+} + {\mathcal H}(-\sigma)G_{\mu\nu}^{-}
- \delta(\sigma) e_{\mu}^{a}e_{\nu}^{b}\left([K_{ab}]-g_{ab}[K]\right),
\end{equation}
where $K_{ab}=e_{a}^{\alpha}e_{b}^{\beta} \nabla_{\beta}n_{\alpha}$ is the extrinsic
curvature and $e_{a}^{\mu}$ is a projection vector onto the hypersurface defined by
$e_{a}^{\alpha} = \partial x^{\alpha}/\partial x^{a}$.
Generically, assuming the energy-momentum tensor in the whole regions {$T_{\mu\nu}
= {\mathcal H}(\sigma)T^{+}_{\mu\nu}+{\mathcal H}(-\sigma)T^{-}_{\mu\nu}
+ \delta(\sigma){\mathcal S}_{\mu\nu}$}, then we have full equations of motion,
\begin{eqnarray}
G_{\mu\nu}^{\pm}={\kappa^2} T^{\pm}_{\mu\nu}, && (\rm exterior~and~interior~regions)\\
{\kappa^2} {\mathcal S}_{ab} = -([K_{ab}]-g_{ab}[K]), &&(\rm on~the~hypersurface) \label{eq:5-1}
\end{eqnarray}
{where ${\kappa^2} = 8\pi G_{N}$.}
Note that Eq. (\ref{eq:5-1}) will determine the dynamical motion of the hypersurface.
Here, the exterior and interior geometries are assumed to be vacuum (or other possible)
geometries and the dynamical collapsing motion of the matter shell (hypersurface) will
produce the final geometry of the exterior spacetime -- this is the usual process of the
gravitational shell-collapse. More precisely, the shell's
equation of motion can be reexpressed in terms of a point-like particles' motion under the
effective potential.
The collapse scenarios of the shell can be analyzed by investigating the behavior of the
effective
potential\footnote{See \cite{Mann:2006yu,Mann:2008rx} for more detailed study of this
formalism and analysis.}.

On the other hand, when we consider the geometrical structure at the region with strong
gravitational fields such
as black hole horizon or curvature singularity, it is expected that the noncommutativity of
spacetimes can remove the point-source structure due to the noncommutative effect of $\Theta$,
where $[x^{\mu},x^{\nu}]=i\Theta^{\mu\nu}$ \cite{witten} with
$\Theta^{\mu\nu} = \Theta~{\rm diag}(\epsilon_{1},\cdots,\epsilon_{D/2})$ \cite{noncom},
where $\Theta$ and $D$ are a constant and the spacetime dimension \cite{noncom2,Kim:2008vi}.
Inspired by this philosophy, a simple mathematics tells us that the delta
functional point source can be replaced by a smeared Gaussian distribution of the
width $\sqrt{\Theta}$, which implies that the energy density of the gravitational
source can be chosen as
\begin{equation}
\label{eq:rhotheta}
\rho_{\Theta} = \frac{M}{(4\pi \Theta)^{3/2}} e^{-\frac{R^2}{4\Theta}}.
\end{equation}
The energy density and the conservation law for a static metric solution with a spherical
symmetry tell
us that the corresponding energy-momentum tensor is given by
$T^{\mu}_{~\nu}={\rm diag} (-\rho_{\Theta},p_{R},p_{\perp},p_{\perp})$,
where the radial and the tangential components of the pressure are
$p_{R} = -\rho_{\Theta}$ and $p_{\perp} = -\rho_{\Theta}
- \frac{1}{2}R\partial_{R}\rho_{\Theta}$.
{Note that in Eq. (\ref{eq:rhotheta}) the energy-momentum tensor vanishes
as $R \rightarrow \infty$
or  $\Theta \rightarrow 0$, which corresponds to the commutative limit.
In the section 4, the black hole formation under the noncommutative
effect $\Theta$ will be investigated. }

\subsection{Schwarzschild Black Hole from the Shell Collapse}\label{sec:SS}

In this section we first consider the Schwarzschild black hole formation from collapsing
shells with the polytropic equation of state, including the perfect fluid and the
Chaplygin gas.
We assume that the metrics in both regions, ${\mathcal V}_{+}$
(outside the shell) and
${\mathcal V}_{-}$ (inside the shell) are given by
\begin{equation}
  \label{eq:1}
  (ds)^2_{{\mathcal V}_{\pm}} = -F_{\pm}(R) dT^2 + \frac{dR^2}{F_{\pm}(R)} + R^2(d\theta^2
  + \sin^2\theta d\phi^2),
\end{equation}
where $F_{+}$ and $F_{-}$ are exterior and interior metrics can be assumed to be a vacuum
solution.
The most general solution is given by the Schwarzschild black
hole one, $F_{\pm} = 1-2M_{\pm}/R$. If $M_{\pm}=0$
it describes a flat space.
Generically, there exist nine possible scenarios depending upon the value
of the parameter $M_{\pm}$ inside and outside the shell. We see that $M_{\pm}>0$ describes
black holes
inside and outside the shell while $M_{\pm} <0$ describes negative mass black holes in both
sides
\cite{Mann:1997jb}. However, the possibility of negative mass black holes will be discussed
later.

Provided we employ a coordinate system   ($t$, $\theta$, $\phi$) on $\Sigma$ at
$R={\mathcal R}(t)$,
the induced metric on the shell becomes
\begin{eqnarray}
  \label{eq:2}
  (ds)_{\Sigma}^2 &=& - F_{\pm} dT^2 + \frac{d{\mathcal R}^2}{F_{\pm}} +
  {\mathcal R}(t)^2 (d\theta^2 + \sin^2\theta d\phi^2)\nonumber\\
  &=& - dt^2 + r_{0}^2 a^2(t)(d\theta^2+\sin^2\theta d\phi^2),
\end{eqnarray}
Continuity of the metric implies that $[g_{ij}]=0$ or
 $F_{\pm}^2 \dot{T}^2 = \dot{\mathcal R}^2 + F_{\pm}$
and ${\mathcal R}^2 = r_{0}^2 a(t)^2$ . However there exists a
discontinuity in the extrinsic curvature of the shell, $[K_{ij}]\ne
0$, since  nonvanishing surface stress-energy  tensor exists. The
extrinsic curvatures on ${\mathcal V}_{\pm}$ are
\begin{equation}
  \label{eq:3}
  K_{~t}^{t~{\pm}} = \frac{d}{d\mathcal R} \sqrt{\left(\frac{d{{\mathcal
  R}}}{dt}\right)^2 + F_{\pm}}, ~~K_{~\theta}^{\theta~{\pm}} =K_{~\phi}^{\phi~{\pm}}=
  \frac{1}{\mathcal R} \sqrt{\left(\frac{d{\mathcal R}}{dt}\right)^2 + F_{\pm}}.
\end{equation}
The surface stress-energy tensor is defined by
\begin{equation}
  \label{eq:4}
  \kappa^2 {\mathcal S}_{ab} = - \left([K_{ab}]-[K]h_{ab}\right),
\end{equation}
where $h_{ab}$ is the induced metric on $\Sigma$.
 The surface stress-energy tensor for a fluid of energy
density $\rho$ and pressure $p$ is assumed to be
\begin{equation}
{\mathcal S}^{ab} = \rho u^{a}u^{b} + p h^{ab} ,\label{eq:1.5}
\end{equation}
where $h_{ab}=g_{ab}+u_a u_b$ is an induced metric on $\Sigma$, and $u^{a}$ is the
shell's velocity.
The surface stress-energy tensor can be straightforwardly evaluated
\begin{equation}
  \label{eq:5}
  {\kappa^2}{\mathcal S}^{t}_{~t} = \frac{2}{{\mathcal R}}
  (\beta_{+}-\beta_{-}),~~\kappa^2 {\mathcal S}^{\theta}_{~\theta} = \frac{d}{
  d{\mathcal R}}({\beta}_{+} - {\beta}_{-}) + \frac{1}{\mathcal R} (\beta_{+}-\beta_{-}),
\end{equation}
where $\beta_{\pm} \equiv \sqrt{\dot{\mathcal R}^2 + F_{\pm}}$. Using
Eqs. (\ref{eq:1.5}) and (\ref{eq:5}), we have two relations,
\begin{eqnarray}
  &&(\beta_{+}-\beta_{-}) + \frac{\kappa^2}{2} \rho {\mathcal R}=0 \label{eq:6}\\
  &&\frac{d}{d{\mathcal R}}({\beta}_{+} - {\beta}_{-}) + \frac{1}{\mathcal R}(\beta_{+}-\beta_{-}) - \kappa^2 p = 0.\label{eq:7}
\end{eqnarray}
These two relations can be easily reduced to a differential equation for the energy density
\begin{equation} \label{eq:9}
\frac{d \rho}{d \log {\mathcal R}} + 2 (\rho +  p) = 0.
\end{equation}
For the polytropic-type matter, the equation of state is given by
\begin{equation}
\label{eq:eos}
p = \omega \rho^{(n+1)/n} ,
\end{equation}
where $\omega$ is the equation of state parameter{, which has been introduced and used in
diverse astrophysical situations such as the Lane-Emden models \cite{chandra}.
In particular, Eq. (\ref{eq:eos}) can be reparameterized as $p=c \theta^{n+1}$ and
$\rho = \lambda  \theta^n$, where $c$ and $\lambda$ are central pressure and central
energy density, respectively, and $n$ denotes the polytropic index. Many stellar models
are led by some specific polytropic indices - neutron stars between $n=0.5$ and $n=1$,
degenerate star cores of white dwarfs, brown dwarfs, and giant gaseous planets like
Jupiter for $n=1.5$, and main sequence stars such as Sun for $n=3$.}
Moreover it can also describe a Chaplygin gas fluid by choosing $n=-1/(\nu+1)$ and
$\omega = -A$ where $A>0$ \cite{Mann:2008rx,kmp}.\footnote{{In three-dimensional
theory \cite{Cornish:1991kj},
the polytropic index $n$ describes many sorts of known fluids -- for example, it
describes constant energy density for $n=0$, nonrelativistic degenerate fermions
for $n=1$, nonrelativistic matter or radiation pressure for $n=2$, linear (perfect)
fluid for $n\rightarrow \infty$.}}

The energy density $\rho$ {{assumed to be positive}} is given by a function of ${\mathcal R}$
by solving Eq. (\ref{eq:9})
\begin{eqnarray}
\label{eq:rho}
\rho &=& {\left[ (\omega + \rho_0^{-1/n}) \left( \frac{\mathcal R}{{\mathcal R}_0}
\right)^{2/n} - \omega \right]^{-n}} ~~~~~~~~({\rm finite}~n)\\
\rho &=& \rho_0\left(\frac{{\mathcal R}_{0}}{{\mathcal R}}\right)^{2\omega+2}.
\qquad\qquad\qquad\qquad\qquad({\rm perfect~fluid})
\end{eqnarray}
Note that for finite and positive $n$'s, the energy density diverges at
\begin{equation}
{\mathcal R}_{s}\equiv {\mathcal R}_0\left(\frac{\omega}{\omega+\rho_0^{-1/n}}\right)^{n/2}.
\end{equation}
Moreover, for even $n$'s, the density always has positive values while for odd $n$'s, they
are negative when ${\mathcal R} < {\mathcal R}_s$ and positive when ${\mathcal R}
> {\mathcal R}_s$, respectively.
In this relation, ${\mathcal R_0}$ is an arbitrary position of the shell and $\rho_0$ is
the energy density of the matter
on the shell located at ${\mathcal R}={\mathcal R_0}$.

Combining Eqs. (\ref{eq:6}) and (\ref{eq:7}) leads to an equation of motion for the shell's
edge
as
\begin{equation}
\label{eq:eqnmotion}
\dot{\mathcal R}^2 + V_{eff}({\mathcal R})= 0 ,
\end{equation}
where the effective potential $V_{eff}$ is given by
\begin{equation}
V_{eff}({\mathcal R}) = \frac{1}{2} (F_+ + F_-) -  \frac{(F_+ - F_-)^2}{(\kappa^2 \rho
{\mathcal R})^2}
- \frac{1}{16} (\kappa^2 \rho {\mathcal R})^2 .
\end{equation}
By defining $x\equiv {\mathcal R}/{\mathcal R}_0$, $\tau \equiv t/{\mathcal R}_0$, and
using Eq. (\ref{eq:rho}), the effective potential can be simplified to
\begin{equation}
\dot{x}^2 + V_{eff}(x) = 0,
\end{equation}
where the effective potential for finite $n$'s is
\begin{equation}
\label{eq:pot1}
V_{eff}(x) = 1 - \frac{\mu_{+}}{x} - \frac{\mu_{-}^2 \varrho_0^2}{4x^4} (x^{2/n}-c_n)^{2n}
- \frac{\varrho_0^2 x^2}{(x^{2/n}-c_n)^{2n}},
\end{equation}
with
\begin{equation}
\mu_{\pm} = \frac{M_{+}\pm M_{-}}{{\mathcal R}_0}, ~~c_{n} = \frac{\omega}{\omega
+ \rho_0^{-1/n}},~~\varrho_0 = \frac{4 (\omega+\rho_0^{-1/n})^{n}}{\kappa^2 {\mathcal R}_0}.
\end{equation}
For perfect fluids, it is given by
\begin{equation}
\label{eq:pot2}
V_{eff}(x) = 1 - \frac{\mu_{+}}{x} - \frac{\bar{\varrho}_{0}^2}{x^{2+4\omega}}
- \frac{\mu_{-}^2 x^{4\omega}}{4\bar{\varrho}_{0}^2},
\end{equation}
where
\begin{equation}
\bar{\varrho}_{0}^2 \equiv \frac{1}{16} \kappa^4 \rho_0^2 {\mathcal R}_{0}^{2-8\omega}.
\end{equation}

It is easy to check that the effective potential negatively diverges at
$x=x_{s}\equiv c_n^{n/2}$ and $x=0$ ($c_n>0$) or only at $x=0$ ($c_n<0$)
\footnote{For $c_n=0$, we shall not discuss this case here since it describes the case
of dust shell ($\omega=0$).}. When $x \to \infty$ it approaches $V_{eff} (x)
\rightarrow v_f \equiv 1-\mu_{-}^2\varrho_0^2/4$ for finite $n$'s,
which is depicted in Fig. \ref{fig:VeffSS}.
Whereas for an infinite $n$ (perfect fluid case),
the behaviors of the potential are classified by four options depending on the value of
the equation of state parameter $\omega$, which is shown in Fig. \ref{fig:sspfshell}.
Note that the energy density diverges at $x=x_{s}$ and $x=0$ ($x=0$ only) leads to the
divergence of the effective potential at those points for finite $n$
(infinite $n$). So the intrinsic Ricci scalar on the shell
\begin{equation}
{\sf R}_{\mu}^{\mu}[\Sigma] = \frac{2}{x^2} \left[ 2x\ddot{x}  + \dot{x}^2
+ \frac{1}{{\mathcal R}_0^2}\right] = \frac{2}{x^2} \left[ -x \frac{d }{dx}V_{eff}(x)
-V_{eff}(x) + \frac{1}{{\mathcal R}_{0}^2}\right]
\end{equation}
is also singular at the same points.


\fig{8.5cm}{8.0cm}{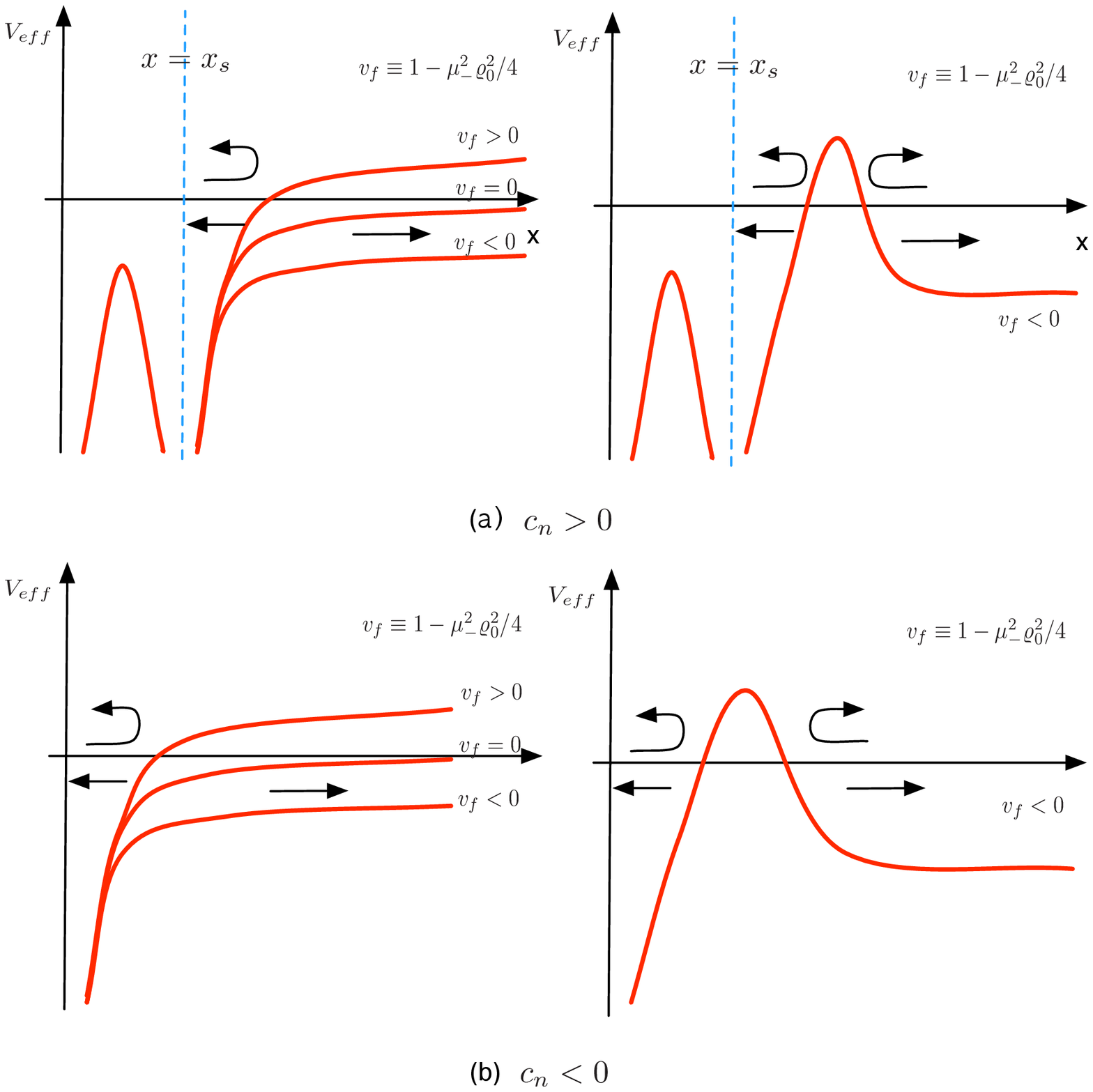}{\small A Cartoon view of the effective potential for
collapsing polytropic shell with finite $n$'s to either the Schwarzschild black hole
to a naked singularity.}{fig:VeffSS}

\fig{10cm}{8cm}{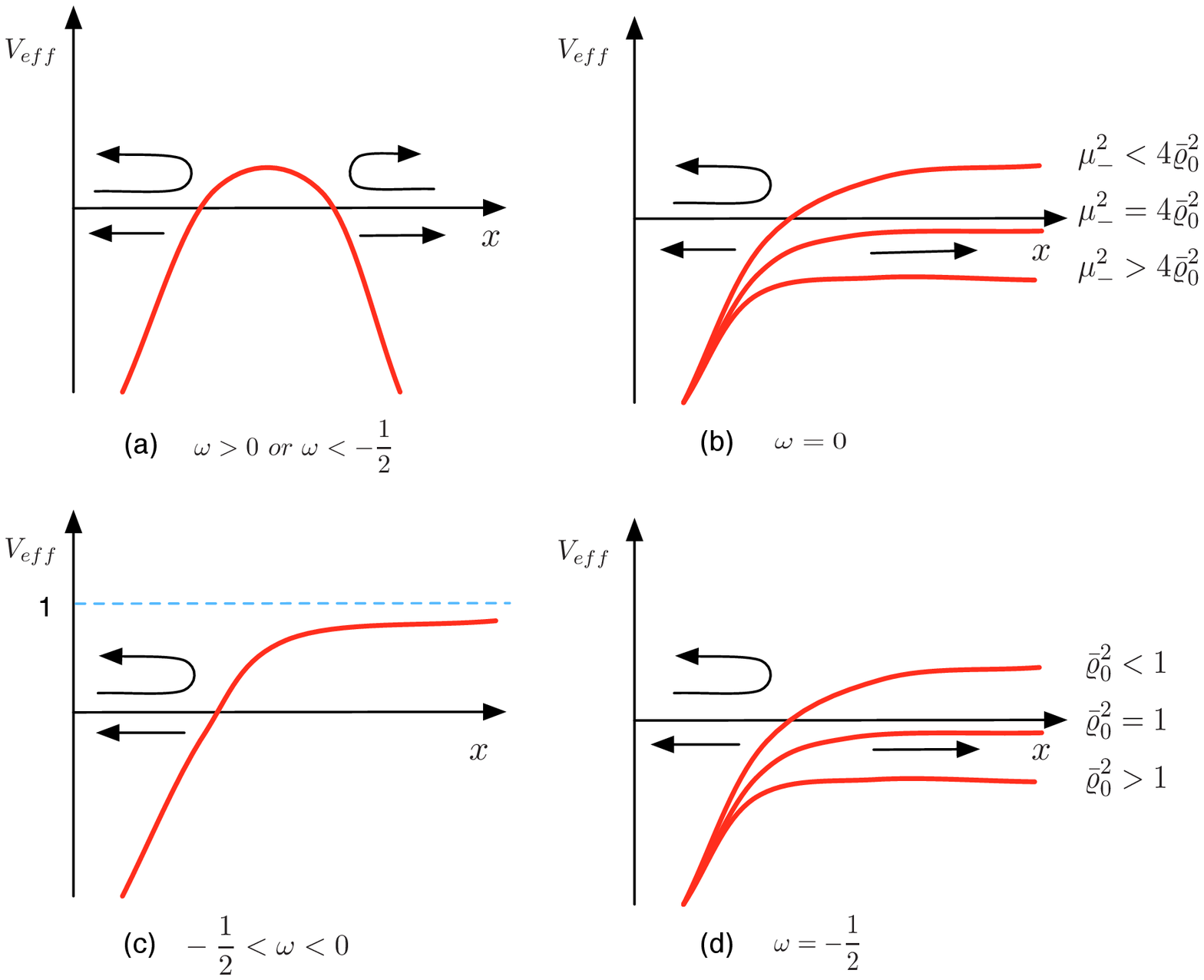}{\small A Cartoon view of the effective potential for
collapsing perfect fluid shell to either the Schwarzschild black hole to a naked
singularity.}{fig:sspfshell}


For perfect fluids, there are four options in the collapse scenario, depending on the
equation-of-state parameter and shell's initial data. First,
as seen in Fig. \ref{fig:sspfshell} (a), for $\omega > 0$ or $\omega < -1/2$,
the effective potential is convex up, so the collapse scenario has three choices:
(i) there are two roots -- if the shell's initial position is in the left-hand side,
it always collapses to a point at $x=0$, creating a curvature singularity while the
shell's position is in the right-hand side, it expands indefinitely,
(ii) there is one degenerate root -- the shell is in the metastable root, which leads to
either collapse or expansion, (iii) there are no roots -- the shell either collapses to a
point or expands indefinitely, depending on its initial data.
In above all cases, if we assume the black hole in the exterior spacetime and flat interior
spacetimes, the black hole can be formed by collapsing perfect fluid shells.

For $-1/2<\omega <0$, the effective potential negatively diverges at $x=0$ while it diverges
positively at $x\rightarrow \infty$, which implies that the shell always collapses to a point
regardless of its initial condition -- i.e., the black hole can be formed all the time,
regardless of its initial data. For the values of $\omega=0$ and $\omega=-1/2$, the behavior
of the potential has the same pattern in that it diverges at $x=0$ and approaches a finite
value at asymptotic region. Therefore, if there is one root, the shell always collapses to
a point regardless of its initial condition while if not, the shell either collapses or
expands indefinitely, depending on the initial data, which implies the critical phenomenon
of the black hole formation in that this behavior depends on the magnitude between black hole
masses and the shell's initial masses.

On the other hand, for the case of finite $n$'s, one of interesting features is that the
singularity at $x=0$ is shifted to finite position of $x=x_{s}$, so the spherical singular
shell is formed after the collapse when $c_{n}>0$ while for $c_{n}<0$, the behavior of the
potential is similar to Fig. \ref{fig:sspfshell} (b) and (d).


As seen in Eq. (\ref{eq:eqnmotion}), the equation of motion is highly nonlinear to obtain an
exact solution but its numerical solution can be easily obtained since it is
the first-order differential equation however the effective potential is extremely complicated.
Some numerical behaviors for a few parameter sets are shown in Fig. \ref{fig:numsol}
by using the fourth order Runge-Kutta method.

\figs{6cm}{6cm}{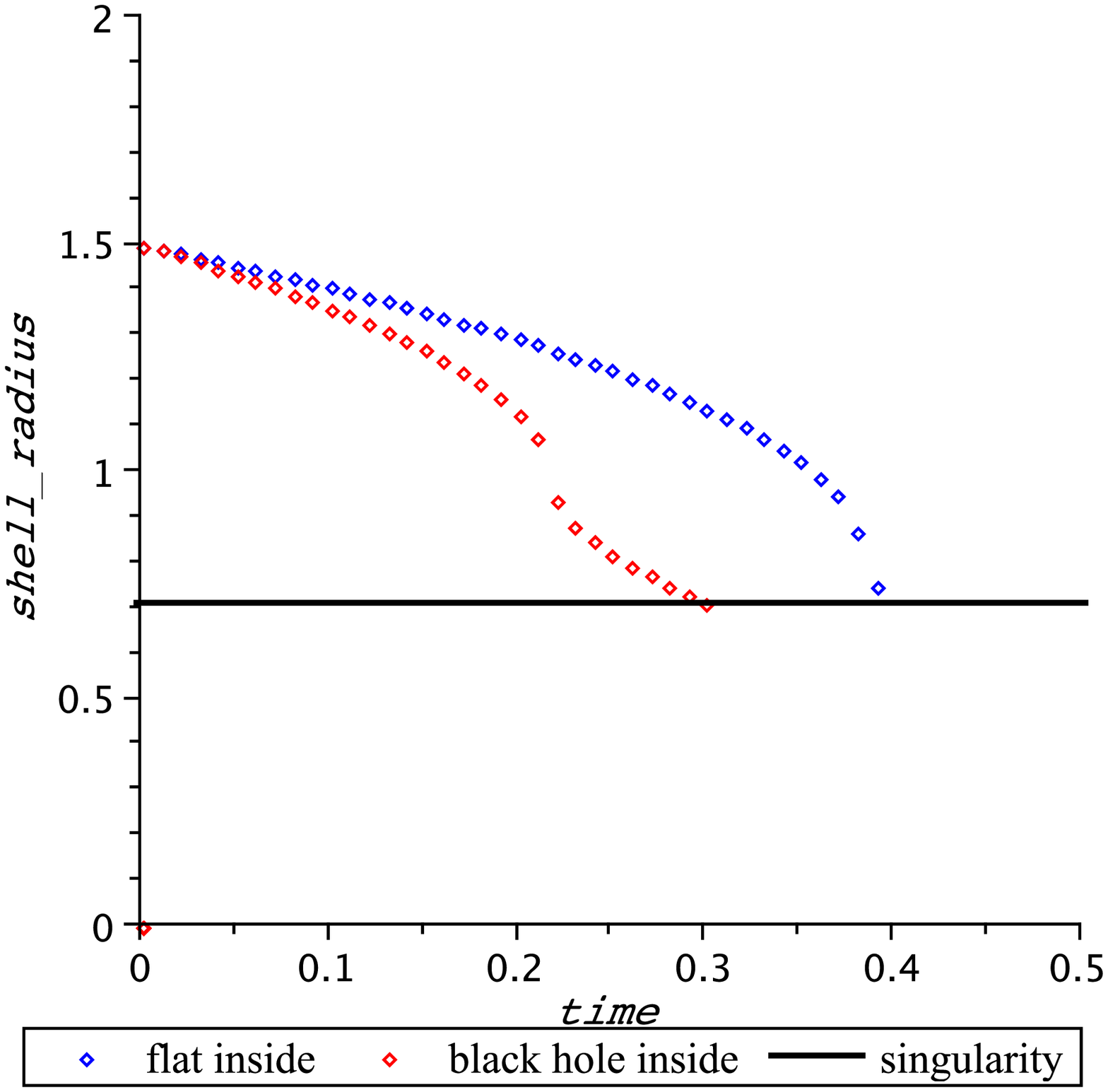}{6cm}{6cm}{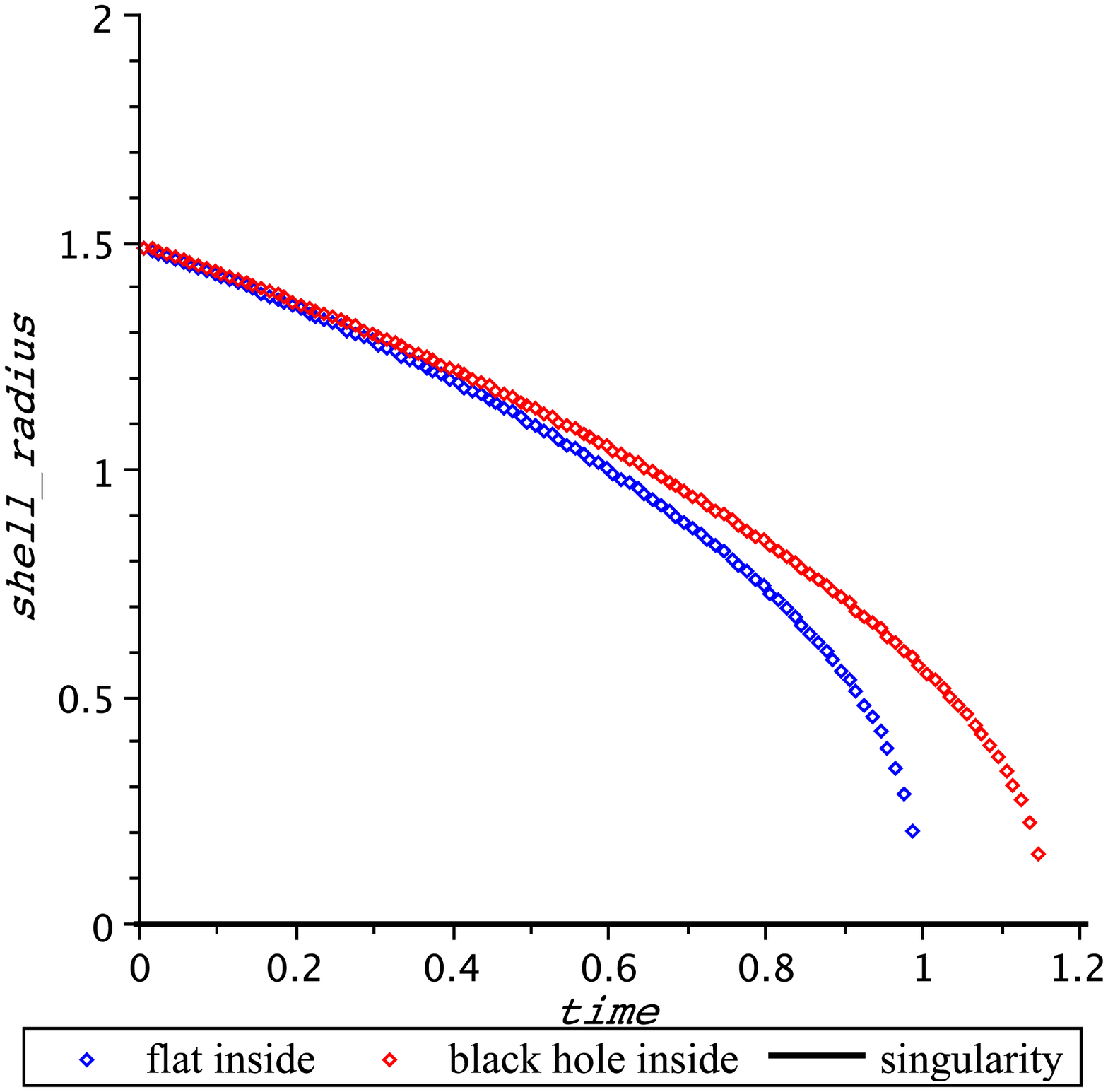}{\small Some behaviors of numerical
solutions of collapsing shells (LHS: $c_n=1$ \& RHS: $c_n=-1$) and specific parameters
are set to be $M_{+}=1$, $\rho_0=1$, $n=1$. The red dot-line is for the interior black hole
($M_{-}=0.5$) while the blue dot-line is for the interior flat geometries
($M_{-}=0.0$).}{fig:numsol}

The previous analysis for the Chaplygin gas shell also shows the similar behaviors as
studied before. The crucial point of this collapsing behavior is whether or not the matter
is pressureless. As shown in three-dimensional case \cite{Mann:2006yu,Mann:2008rx}, the
pressure of collapsing matter can stretch the singularity to finite size since the curvature
and the energy density on the shell diverges at that point. This singularity can be screened
by an event horizon for an appropriate initial data while it is equivalently possible to be
exposed outside the geometry, depending on initial condition, which might require a certain
modification of formalism beyond the classical {theory of} gravity.

\section{Gravitational Collapse of the Smeared Source}\label{sec:NCSS}

In this section we consider the noncommutative Schwarzschild black hole formation from
collapsing shells,
including the new matter related to the noncommutativity
as well as the polytropic matter on the shell.
This is the generalization of the shell collapsing on
the noncommutative background. Notice that to solve the junction equation we need new matter
different with
the polytropic matter on the shell, whose energy density depending on the noncommutative
parameter $\Theta$
and the shell position $R$ will be calculated.
Another interesting point of this section is to check weather the noncommutative effect
of the background
can get rid of the naked singular shell.

By introducing the smeared energy density (\ref{eq:rhotheta}) to the bulk, we assume that the
metrics in both regions, ${\mathcal V}_{+}$
(outside the shell) and
${\mathcal V}_{-}$ (inside the shell) are given by
\begin{equation}
  \label{eq:1}
  (ds)^2_{{\mathcal V}_{\pm}} = -F_{\pm}(R) dT^2 + \frac{dR^2}{F_{\pm}(R)} + R^2(d\theta^2
  + \sin^2\theta d\phi^2),
\end{equation}
where $F_{+}$ and $F_{-}$ are exterior and interior metrics that can be assumed to be an
appropriate solution of the Einstein equation, so we set
\begin{equation}
\label{eq:F}
F_{\pm}(R) = 1- \frac{2m_{\pm}(R)}{R} = 1- \frac{4M_{\pm}}{R\sqrt{\pi}}
\gamma\left(\frac{3}{2},\frac{R^2}{4\Theta}\right),
\end{equation}
where $\gamma(3/2,x^2)$ is an incomplete gamma function\footnote{See Ref. \cite{as} for
more details on the gamma function and its properties.} defined as
\begin{eqnarray}
&&\gamma\left(\frac{1}{2},x^2\right) \equiv 2 \int_{0}^{x} e^{-t^2} dt = \sqrt{\pi}
{\rm erf} (x)\nonumber\\
&&\gamma(a+1,x) = a\gamma(a,x)-x^{a} e^{-x}.
\end{eqnarray}

{Note that the metric (\ref{eq:F}) becomes the standard Schwarzschild metric in the limit of
$R/\sqrt{\Theta} \rightarrow \infty$ (commutative limit). On the other hand, the surface
energy-momentum
tensor (\ref{eq:1.5}) should be modified by
$\rho \rightarrow \rho_{s} + \rho_{\Theta}$ and $p \rightarrow p_{s} + p_{\perp}$,
where} {$\rho_s$ and $p_s$ are the energy density and pressure of the polytropic-type
matter in the previous section.}
Then, the junction equation becomes
\begin{eqnarray}
  &&(\beta_{+}-\beta_{-}) + \frac{\kappa^2}{2}\rho {\mathcal R}=0 \label{eq:4.3}\\
  &&\frac{d}{d{\mathcal R}}({\beta}_{+} - {\beta}_{-}) + \frac{1}{\mathcal R}(\beta_{+}
  -\beta_{-})
  - \kappa^2 p = 0 .
  \label{eq:4.4}
\end{eqnarray}
The equation for the energy density related to the noncommutativity  is reduced to
\begin{equation}
 \frac{d \rho_{\Theta}}{d \log {\cal R}} + 2 (\rho_{\Theta} + p_{\perp}) = 0.
\end{equation}
Using the fact that the pressure $p_{\perp}$ can be rewritten as
\begin{equation}
p_{\perp} = - \left( 1 - \frac{{\mathcal R}^2}{4 \Theta} \right) \rho_{\Theta} ,
\end{equation}
the energy density becomes
\begin{equation}
\rho_{\Theta} =   \bar{\rho} \ e^{-\frac{{\mathcal R}^2-{\mathcal R}_0^2}{\Theta}},
\end{equation}
where $\bar{\rho}$ is the value at ${\mathcal R} ={\mathcal R}_0$.
Note that the above energy density is different with that of the polytropic matter, which
means that
for describing the shell collapse in the noncommutative background consistently we should
introduce
new matter smeared in the hypersurface. Interestingly, when ${\mathcal R}/\sqrt{\Theta}
\to 0$, this
matter behaves like the cosmological constant.
Plugging this
into Eq. (\ref{eq:4.3}) yields the equation describing the shell's motion
\begin{equation}
\label{eq:motionV}
\dot{\mathcal R}^2 + V_{eff}({\mathcal R})= 0 ,
\end{equation}
with the effective potential
\begin{equation}
\label{eq:PotV}
V_{eff}({\mathcal R}) = \frac{1}{2} (F_+ + F_-) - \frac{(F_+ - F_-)^2}{(\kappa^2
{\mathcal R})^2}
\rho^{-2}
- \frac{1}{16} (\kappa^2  {\mathcal R})^2 \rho ^2 ,
\end{equation}
where $\rho$ is given by
\begin{equation}
\rho = {\left[ (\omega + \rho_0^{-1/n}) \left( \frac{\mathcal R}{{\mathcal R}_0}
\right)^{2/n} - \omega \right]^{-n}}
+  \bar{\rho}\ e^{-\frac{{\mathcal R}^2-{\mathcal R}_0^2}{\Theta}} .
\end{equation}
In the case of $n \to \infty$, the energy density becomes
\begin{equation}
\rho = \rho_0
+  \bar{\rho} \ e^{-\frac{{\mathcal R}^2-{\mathcal R}_0^2}{\Theta}} .
\end{equation}
Here if we set ${\mathcal R}/{\mathcal R}_{0} \equiv x$ and $t/{\mathcal R}_{0} \equiv
\tau$ as before, then one finds the energy density
\begin{equation}
\label{eq:rhoTheta}
\rho = \left[ (\omega + \rho_{0}^{-1/n})x^{2/n} - \omega\right]^{-n} + \bar{\rho}
e^{-\frac{{\mathcal R}_0^2}{\Theta} (x^2-1)}
\end{equation}
and the behavior of the effective potential with Eqs. (\ref{eq:F}) and (\ref{eq:rhoTheta})
is similar but
slightly modified by the noncommutative effect of $\Theta$.
For the generic polytropic matters with $c_n>0$, there exists a singular point at
$x = x_{s} \equiv \omega^{n/2}/(\omega+\rho_0^{-1/n})^{n/2}$ while it is not the case
for the perfect
fluid of $\rho_{s}$.
{Due to the finiteness of the second term in Eq. (\ref{eq:rhoTheta}),
the noncommutative correction does not affect the typical shape of the effective potential
coming from
the polytropic matter.
In other words, the noncommutative energy density cannot remove the singularity caused by
the polytropic matters.
}
The effective potential for some specific parameters and difference due to the noncommutative
effect are depicted in Fig. \ref{fig:ncpot}.

\figs{6cm}{6cm}{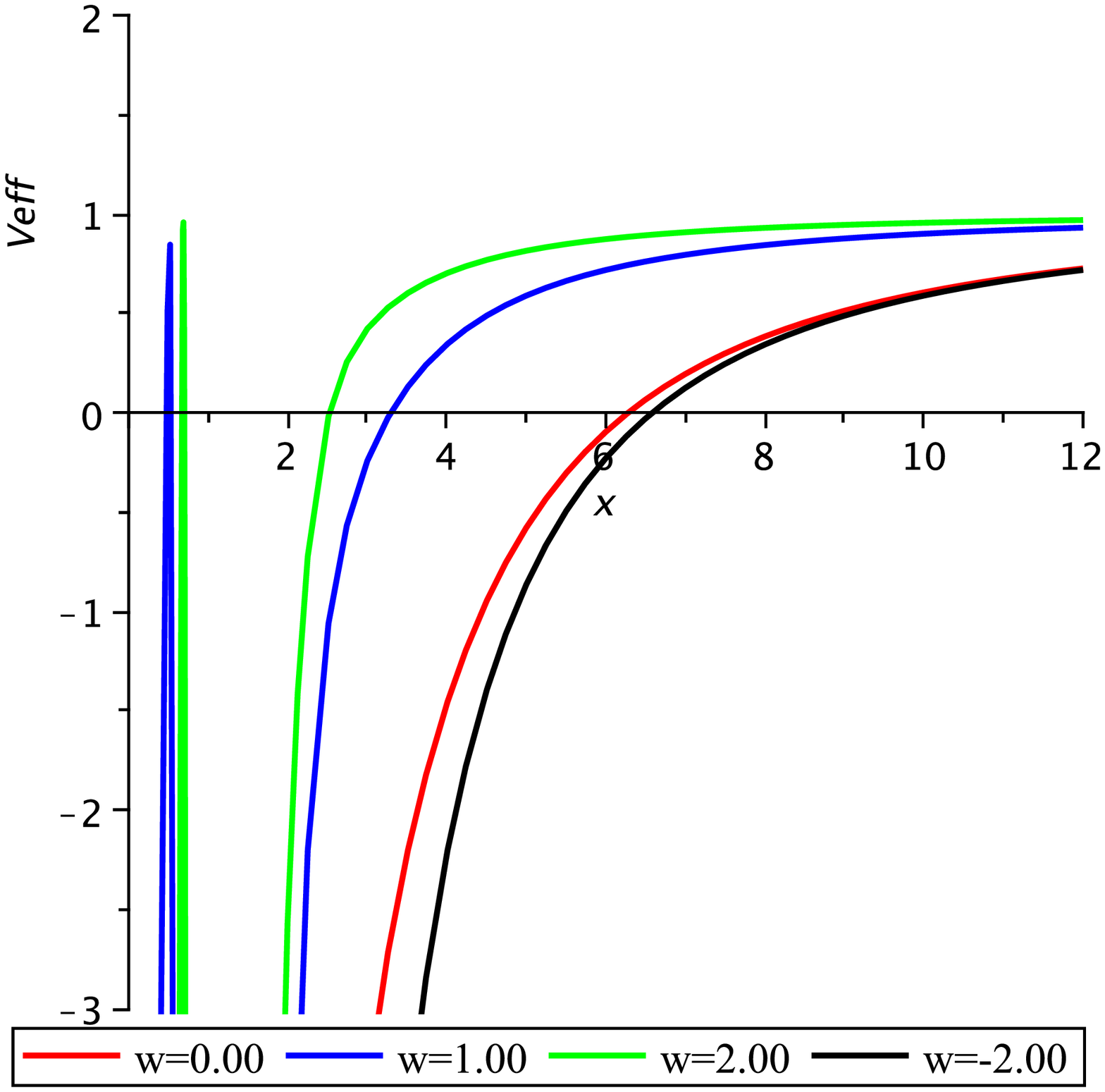}{6cm}{6cm}{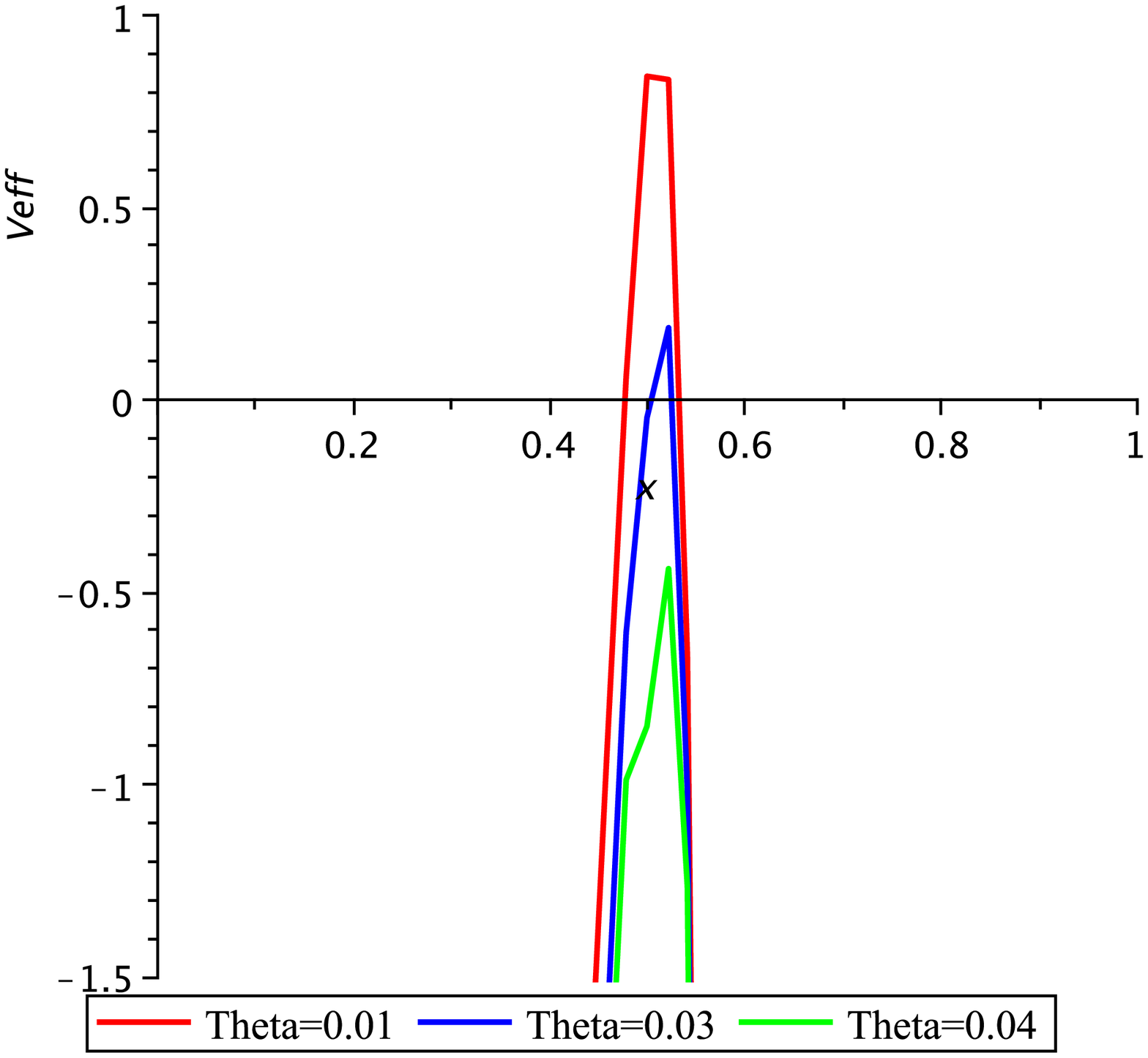}{\small Some behaviors of the
effective potential in varying $\omega$ (LHS). The RHS is focused view for the inner
convex of the effective potential with respect to the value of $\Theta$.}{fig:ncpot}

The main effect of $\Theta$ parameter (noncommutative effect of collapsing matter) is
shown in RHS of
Fig. \ref{fig:ncpot}. As seen in Fig. \ref{fig:sspfshell} (a),
it is found that the convex of potential in $0 < x < x_{s}$ has always negative value
for some
numerical checks in the commutative case.
{Provided we introduce the $\Theta$-effect, then the potential barrier in this region
appears as shown in RHS of Fig. \ref{fig:ncpot}}. {Anyway, as mentioned previously, the
$\Theta$-effect cannot alter the shell's collapsing
behavior.}
{Though the noncommutative effect does not get rid of the singularity of the effective
potential,
it may be helpful to make a black hole solution before the shell touches the singular
point. To see
this, we first calculate the black hole horizon $x_{h}$} such that
\begin{equation}
1-\frac{4M}{x_{h} R_{0}\sqrt{\pi}}\gamma\left(\frac{3}{2},
\frac{x_{h}^2 R_{0}^2}{4\Theta}\right) = 0,
\end{equation}
in which the horizon depends on $\Theta$.
For example, for some fixed numeric values
(here $G=1$, $M_{+}=1$, $M_{-}=0$, $\rho_0=1$, $\omega=1$, $n=1$),
{the position of the horizon is shifted as
\begin{equation}
x_{h} = \left\{ \begin{array}{lll}
         0.3211830425 & {{\rm for}~ \Theta=0.01}\\
         0.6040244364 & {{\rm for}~\Theta=0.05}\\
         1.2899484640 & {{\rm for}~ \Theta=0.50}.
         \end{array} \right.
\end{equation}}
Since $x_{s} = \omega^{n/2}/(\omega+\rho_0^{-1/n})^{n/2} = 0.7071067810$ for the
parameters we are considering above, {some} collapsing scenarios are possible:

$\bullet$ For the choice of $\Theta =0.01$, since the black hole horizon is only
located between the origin and the potential wall, if the initial shell starts collapsing
in $x_{h}<x_{0}<x_{wall}$, it always forms a noncommutative Schwarzschild geometry.
However, the shell's initial position is outside the wall, the collapsing shell will
always encounter the naked singularity at $x=x_{s}$.

$\bullet$ For the choice of $\Theta =0.05$, since the black hole horizon is located
at $0 < x_{h} < x_{s}$, the shell starting from $x_{h} < x_0 < x_{s}$ either forms a
black hole or encounters a singularity at $x=x_{s}$ -- this picture cannot be allowed
since its initial geometry already includes a singular point in the manifold. Another
option is that the shell's initial position is located at $x_{s} < x_0 < x_{max}$,
for which the shell can form a naked singularity at the final stage of the collapse.

$\bullet$ For the choice of $\Theta = 0.50$ case, the black hole horizon is outside the
singular point at
$x_{s}$, so the shell can form a black hole as the shell collapses. All these scenarios
are depicted in
Fig. \ref{fig:nceff}.

\fig{14.5cm}{5.5cm}{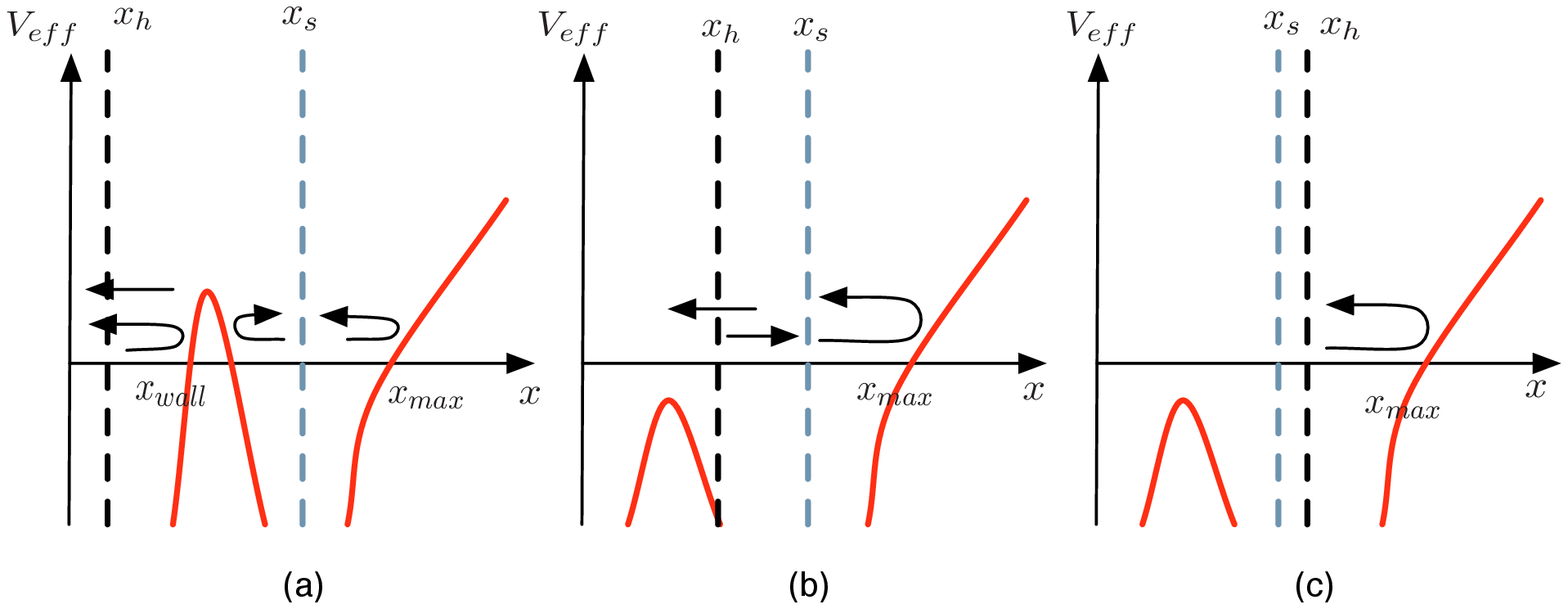}{\small Some collapse scenarios with respect to
noncommutative
parameters and initial data of the shell's position.}{fig:nceff}

{As shown in Fig. \ref{fig:nceff} (c), as the noncommutative effect becomes strong, the black
hole horizon has a large value so that the shell's collapse can make the black hole
before touching the
singular point.}
The equation of motion Eq. (\ref{eq:motionV}) with Eqs. (\ref{eq:PotV}) and
(\ref{eq:rhoTheta}) is highly nonlinear and complicated even if it is simply the
first-order differential equation. Instead of its analytic solution, we can estimate
numerical solutions for some appropriate initial data and fixed parameters.
Fig. \ref{fig:ncnumsol} shows that the collapsing shell of the polytropic matter and
the smeared gravitational source can form a noncommutative Schwarzschild black hole.

\figs{6.5cm}{6.5cm}{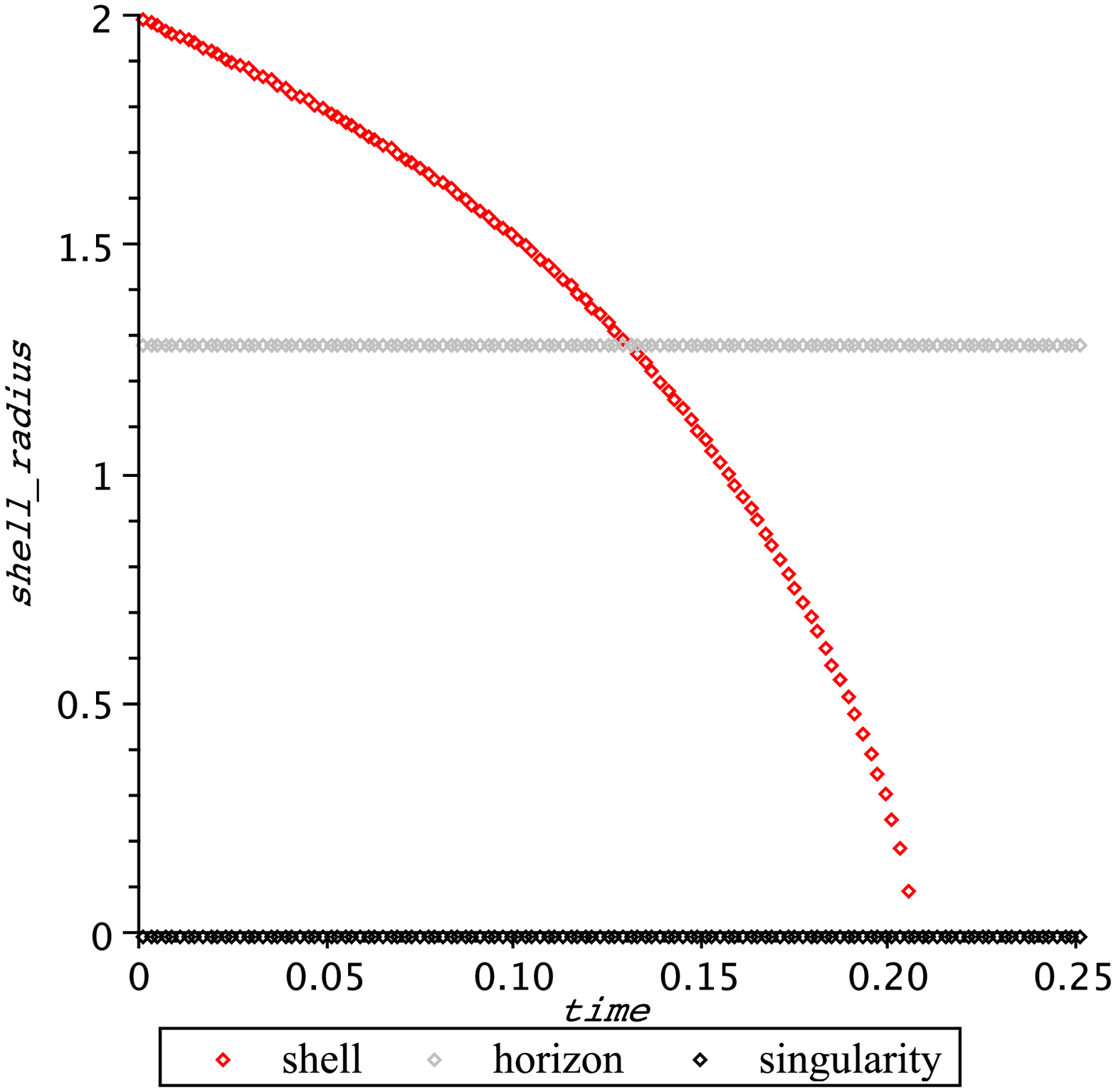}{6.5cm}{6.5cm}{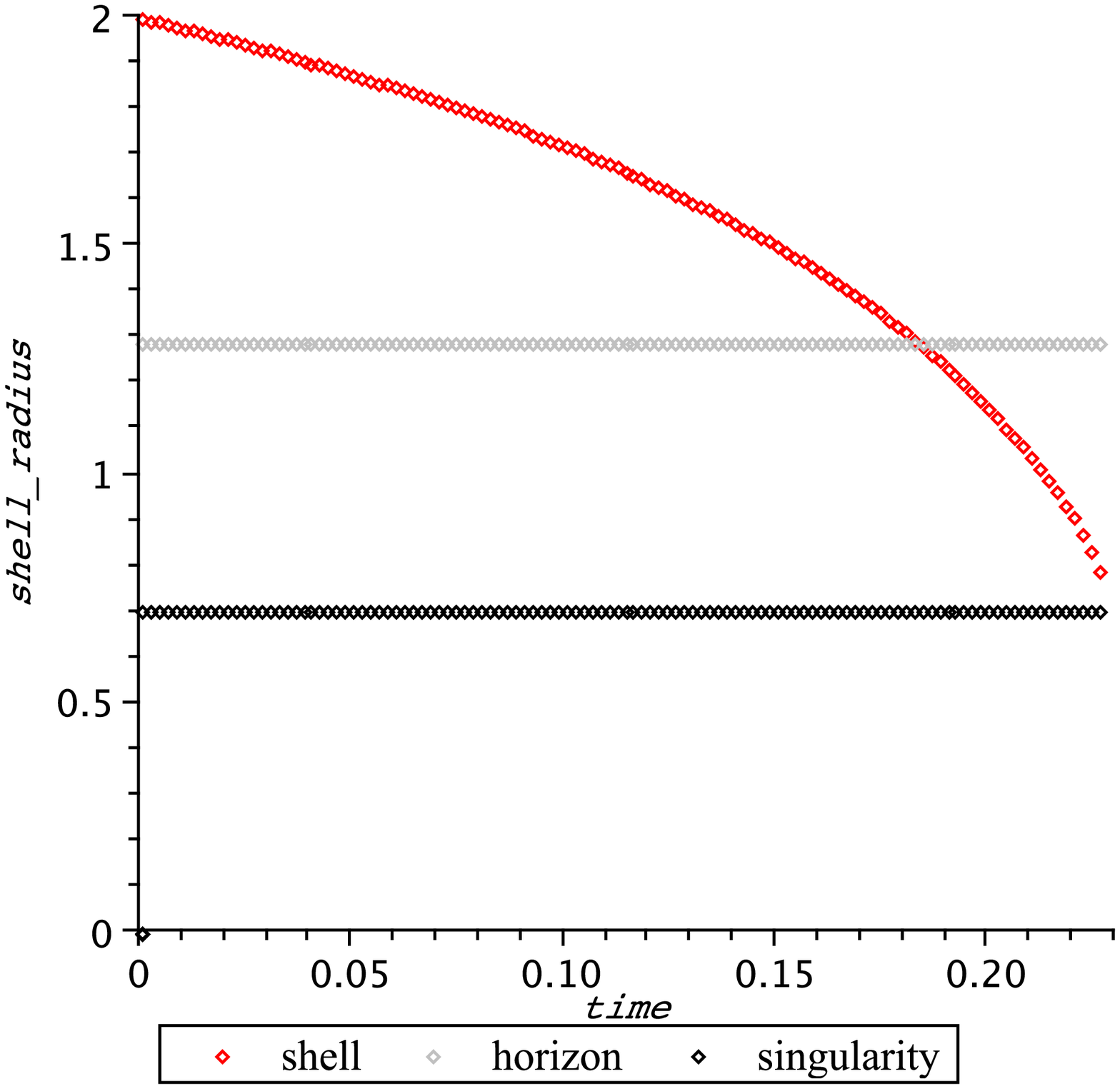}{\small Numerical estimates
of the shell's collapsing solution for the noncommutative black hole formation. The LHS is
for $\omega=0$ (dust) and the RHS is for $\omega=1$ (polytropic pressure). Some parameters
are set by $M_{+}=1$, $M_{-}=0$, $G=1$, $\rho_0=1$, $n=1$, $\Theta=0.5$, and
$R_{0}=1$.}{fig:ncnumsol}

Note that the possibility of a negative energy density that violates the weak energy
condition can be also
took into account. Indeed the study of such configurations has been done in Ref.
\cite{Mann:1997jb},
in which the freely falling dust cloud with a negative energy density collapses to
the black hole with the
negative ADM mass and different topological structure. More precisely, the exterior
topological geometry
is set to be the solution of Einstein equations with cosmological constant $\Lambda <0$
with mass parameter
$M>0$ and topology of $R^2 \times H_{g}$, where $H_{g}$ is a two-surface of genus $g\ge 2$.
Then a
straightforward computation reveals $M_{ADM} = -M(g-1) <0$, which shows that the resulting
geometry
from freely falling dust cloud of negative energy density should be a negative mass black
hole. See Refs.
\cite{ms,Mann:1997jb, Mann:1997iz} for more details on the topological black holes including
the negative
mass black hole.

Concerning the negative energy density \cite{Mann:1997jb}, the aspect seems to be unchanged
comparing to the case of the positive energy density. Indeed, a short glance at the effective
potential in
Eqs. (\ref{eq:pot1}) and (\ref{eq:pot2}) shows that the initial energy density is quadratic,
implying
that the negative energy density does not alter the previous result of the case of $\rho_0>0$.
However, the total energy system must be positive, which implies that
\begin{equation}
E_{\Theta}=\int dV \rho_{\Theta} =\int dV \frac{M}{(4\pi\Theta)^{3/2}}
e^{-\frac{R^2}{4\Theta}} > 0,
\end{equation}
clearly leading $M>0$. The main reason that Ref. \cite{Mann:1997jb} can allow the negative
mass is the only
topological reason of the geometry since the system can produce the positive total energy
with the combination
of the negative mass and the topological genus as mentioned above. For this reason, the
system we are considering
in this paper cannot accept the negative mass case.

\section{Discussions}\label{sec:discuss}

In this paper, we investigated the gravitational shell collapse problem with the smeared
gravitational
source in the context of the {mild noncommutative modification of the space.}
The collapsing matters with the most generic
polytropic equation of state and the Chaplygin gas were considered, which leads to
formation of
either black hole or naked singularity, depending on its initial data. Moreover,
the existence of
the nontrivial pressure yields the stretch of singularity, which produces the singular
shell with
a finite size, regardless of an event horizon. Provided it is cloaked by an event horizon,
the collapsing shell can form a black hole while it forms a naked singular shell if
not -- this
crucially depends on the initial condition. This feature has been already observed in
the lower
dimensional theories of gravity \cite{Mann:2006yu,Mann:2008rx}.
{When considering the noncommutative
effect, we showed that though the noncommutativity cannot remove the singularity caused by
the polytropic matter, it makes the collapsing shell becomes a black hole before touching the
singular point.}

As studied before, the introduction of the smeared gravitational source and the
corresponding slight
modification of the Schwarzschild geometry do not offer a crucial prescription
for the treatment of
naked singularities in the shell collapse formalism. Considering the smeared source
in the presence
of $\Theta$-effect can be regarded as {\it so-called ``semi-noncommutative''} approach
since the
matter part of Einstein's equation has $\Theta$-effect while the shell formalism from
the geometry
parts of Einstein's equation has no such terms. Therefore, in order to deal with this
issue completely,
we, somehow, should modify the gravitational action and have a modified shell formalism.
One of
possible ways of doing this is to add some higher curvature terms to the Einstein-Hilbert
action,
which can alter the junction equation. This scenario can be viable for three-or
five-dimensional
cases because we can add the gravitational Chern-Simons (GCS) terms or Gauss-Bonnet
terms to the
Einstein-Hilbert action. In four dimensions, no viable expression of higher curvature
terms is known
due to the topological reason.
In this sense, a recent suggestion of the Horava-Lifshits gravity
\cite{Horava:2009uw,Horava:2009if}
might draw a great interest since the theory has highly-powered curvature terms
in the action.
Another way to exit the issue is to consider the another alternate theory of
gravitation such as
the scalar tensor theory of Brans-Dicke type, which clearly alters the junction
equations on the shell.
{In these all possibilities,} it is worthwhile to study whether or not the
singularity issue is resolved, which is being
in progress.
However, above all possibilities might be temporal prescriptions of handling a
singularity in that
those corrections only hold in the context of the {\it effective theory}.
The most fundamental
solution to the singularity and the cosmic censorship should be suggested through
the full quantum
theory of gravity.

\vspace{10mm}

{\bf Acknowledgment}
\\
{J. J. Oh would like to thank E. J. Son and S. H. Oh for helpful discussions.
J. J. Oh was supported by the Korea Research Foundation Grant funded by the Korean
Government(MOEHRD, Basic Research Promotion Fund)(KRF-2008-313-C00182). C. Park was
supported by the Korea Science and Engineering Foundation (KOSEF) grant funded by the
Korea government(MEST) through the Center for Quantum Spacetime (CQUeST) of Sogang
University with grant number R11-2005-021.}


\end{document}